# Digital quantum simulation of fermionic models with a superconducting circuit

R. Barends,[1] L. Lamata,[2] J. Kelly,[3] L. García-Álvarez,[2] A. G. Fowler,[1] A. Megrant,[3,4] E. Jeffrey,[1] T. C. White,[3] D. Sank,[1] J. Y. Mutus,[1] B. Campbell,[3] Yu Chen,[1] Z. Chen,[3] B. Chiaro,[3] A. Dunsworth,[3] I.-C. Hoi,[3] C. Neill,[3] P. J. J. O'Malley,[3] C. Quintana,[3] P. Roushan,[1] A. Vainsencher,[3] J. Wenner,[3] E. Solano,[2,5] and John M. Martinis[1,3]

[1]*Google Inc., Santa Barbara, CA 93117, USA*
[2]*Department of Physical Chemistry, University of the Basque Country UPV/EHU, Apartado 644, E-48080 Bilbao, Spain*
[3]*Department of Physics, University of California, Santa Barbara, CA 93106, USA*
[4]*Department of Materials, University of California, Santa Barbara, CA 93106, USA*
[5]*IKERBASQUE, Basque Foundation for Science, Maria Diaz de Haro 3, 48013 Bilbao, Spain.*

**Simulating quantum physics with a device which itself is quantum mechanical, a notion Richard Feynman originated [1], would be an unparallelled computational resource. However, the universal quantum simulation of fermionic systems is daunting due to their particle statistics [2], and Feynman left as an open question whether it could be done, because of the need for non-local control. Here, we implement fermionic interactions with digital techniques [3] in a superconducting circuit. Focusing on the Hubbard model [4, 5], we perform time evolution with constant interactions as well as a dynamic phase transition with up to four fermionic modes encoded in four qubits. The implemented digital approach is universal and allows for the efficient simulation of fermions in arbitrary spatial dimensions. We use in excess of 300 single-qubit and two-qubit gates, and reach global fidelities which are limited by gate errors. This demonstration highlights the feasibility of the digital approach and opens a viable route towards analog-digital quantum simulation of interacting fermions and bosons in large-scale solid state systems.**

The key to simulation is mapping a model Hamiltonian onto a physical system. For fermionic models, a successful approach has been to use physical systems which are natively fermionic. Cold gases of fermionic atoms have performed hallmark experiments in the analog simulation of transport properties and magnetism [6, 7]. However, analog systems are limited to specific classes of problems, and designing the specific interactions is challenging. In digital quantum simulation, interactions can be arbitrarily controlled; already, a local spin Hamiltonian was simulated in ion traps [8]. Yet, the digital approach requires complex sequences of logic gates, especially for non-local control, which hinge on carefully constructed interactions between subsets of qubits in a larger system; a practical obstacle for several platforms. A digital fermionic simulation has therefore remained an open challenge. With recent advances in architecture and control of superconducting qubits [9–11], we can explore universally implementing fermionic models, in a first demonstration of digital quantum simulations in the solid state.

Quantum simulation of fermionic models is highly desirable, as computing the properties of interacting particles is classically difficult. Determining static properties with quantum Monte Carlo techniques is already complicated due to the sign problem [12], arising from anticommutation, and dynamic behaviour is even harder. Here, we use the Jordan-Wigner transformation [13] to map fermionic models onto physical qubits. In this approach, the required number of gates scales only polynomially with the number of modes [5]. Moreover, the model system is not limited to the dimensionality of the physical system, allowing for the simulation of fermionic models in two and three spatial dimensions [5, 14]. Furthermore, bosonic degrees of freedom, discrete [15] or continuous [16], may be also introduced with an analog-digital quantum simulator.

At low temperatures, classes of fermionic systems can be accurately described by the Hubbard model. Here, hopping (strength $V$) and repulsion (strength $U$) compete (see Fig. 1a), capturing the rich physics of many-body interactions such as insulating and conducting phases in metals [17, 18]. The generic Hubbard Hamiltonian is given by: $H = -V \sum_{\langle i,j \rangle} \left( b_i^\dagger b_j + b_j^\dagger b_i \right) + U \sum_{i=1}^{N} n_{i\uparrow} n_{i\downarrow}$, with $b$ the fermionic annihilation operator, and $i, j$ running over all adjacent lattice sites. The first term describes the hopping between sites and the last term the on-site repulsion. It is insightful to look at a fermionic two-mode example,

$$H = -V \left( b_1^\dagger b_2 + b_2^\dagger b_1 \right) + U b_1^\dagger b_1 b_2^\dagger b_2. \quad (1)$$

We can express the fermionic operators in terms of Pauli and ladder operators using the Jordan-Wigner transformation [13]: $b_1^\dagger = I \otimes \sigma^+$ and $b_2^\dagger = \sigma^+ \otimes \sigma_z$, where the $\sigma_z$ term ensures anticommutation. In essence, we use non-local control and map a local fermionic Hamiltonian to a local spin Hamiltonian. The qubits act as spins, and carry the fermionic modes (Fig. 1a-b). A fermionic mode is either occupied or unoccupied, and spinless – the spin degree of freedom is implemented here by using four modes to simulate two sites with two spins. We note that for higher spatial dimensions this approach is still viable, the only difference is that the local fermionic Hamiltonian now maps to a nonlocal spin Hamiltonian, which can be efficiently implemented as recently shown [5, 14]. Using the above transformation, the Hamiltonian becomes

$$H = \frac{V}{2}(\sigma_x \otimes \sigma_x + \sigma_y \otimes \sigma_y) + \frac{U}{4}(\sigma_z \otimes \sigma_z + I \otimes \sigma_z + \sigma_z \otimes I), \quad (2)$$

which can be implemented with separately tunable $\hat{X}\hat{X}$, $\hat{Y}\hat{Y}$, and $\hat{Z}\hat{Z}$ interactions. Here, we use the convention to map an excited fermionic mode $|1\rangle$ (excited logical qubit) onto a qubit's physical groundstate $|g\rangle$, and a vacuum fermionic mode $|0\rangle$ (ground logical qubit) onto a qubit's physical excited state $|e\rangle$.

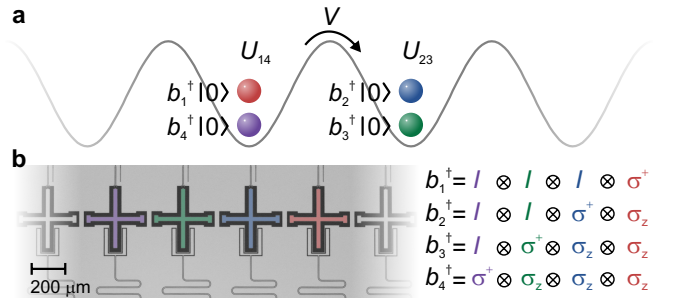

FIG. 1. **Model and device.** (a) Hubbard model picture with two sites and four modes, with hopping strength $V$ and on-site interactions $U$. The creation of one excitation from the groundstate is shown for each mode. (b) Optical micrograph of the device. The coloured cross-shaped structures are the used Xmon transmon qubits. The construction of the fermionic operators for four modes is shown on the right.

Our experiments use a superconducting nine-qubit multipurpose processor, see Fig. 1b. Device details can be found in Ref. [19]. The qubits are the cross-shaped structures [20] patterned out of an aluminium film on a sapphire substrate. They are arranged in a linear chain with nearest-neighbour coupling. Qubits have individual control, using microwave and frequency-detuning pulses (top), and readout is done through dispersive measurement (bottom) [21]. By frequency tuning of the qubits, interactions between adjacent pairs can be separately turned on and off. This system allows for implementing non-local gates, as it has a high level of controllability, and is capable of performing high fidelity gates [9, 22]. Importantly, single- and two-qubit gate fidelities are maintained when scaling the system to larger numbers of qubits.

The basic element used to generate all the interactions is a simple generalization of the controlled-phase (CZ) entangling gate (Fig. 2a-b). We implement a state-dependent frequency pull by holding one qubit steady in frequency and bringing a second qubit close to the avoided level crossing of $|ee\rangle$ and $|gf\rangle$ using an adiabatic trajectory [23]. By tuning this trajectory, we can implement a tunable CZ$_\phi$ gate. During this operation, adjacent qubits are detuned away in frequency to minimize parasitic interactions. The practical range for $\phi$ is 0.5-4.0 rads; below this range parasitic $\hat{Z}\hat{Z}$ interactions with other qubits become relevant, and above this range population starts to leak into higher energy levels (Supplementary Information and Refs. [9, 19]). Using single qubit gates and two entangling gates, we can implement the tunable $\hat{Z}\hat{Z}$ interactions, as shown in Fig. 2c. In this gate construction, the $\pi$-pulses naturally suppress dephasing [24].

First, we have experimentally verified that the encoded fermionic operators anticommute, see Fig. 2d, by implementing the following anticommutation relation $\{b_1, b_2^\dagger\} + \{b_2, b_1^\dagger\} = 0$. The latter can be separated into two non-trivial Hermitian terms: $b_1 b_2^\dagger + b_2 b_1^\dagger$ and $b_1^\dagger b_2 + b_2^\dagger b_1$. Their associated unitary evolution, $U = \exp(-i\frac{\phi}{2}(b_1 b_2^\dagger + b_2 b_1^\dagger))$ for the first one, has been implemented using gates with strength $\phi = \pi$. The measured process matrices ($\chi$) for these terms are determined using quantum process tomography, and constrained to be physical (Supplementary Information). We find that the processes are close to the ideal, with fidelities $\text{Tr}(\chi_{\text{ideal}}\chi) = 0.95, 0.96$. As the Hermitian terms sum up to zero, their unitary evolutions combine to the identity. We find that the sequence of both processes yields in fact the identity, as expected for anticommutation, with a fidelity of 0.91.

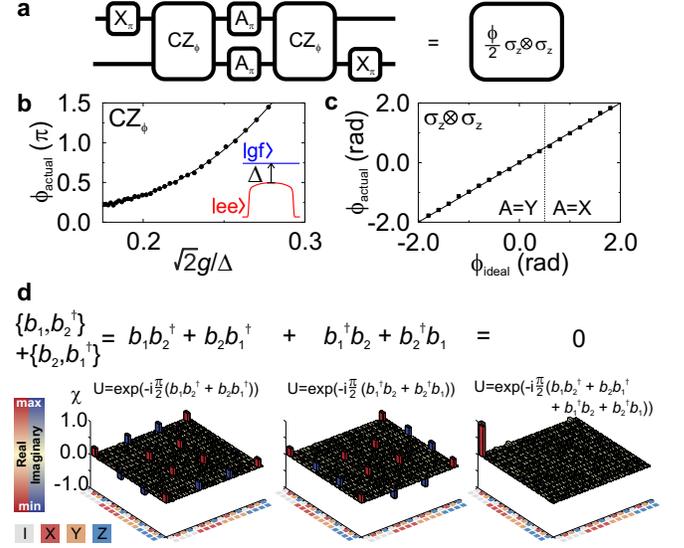

FIG. 2. **Gate construction and operator anticommutation.** (a) Construction of the gate $U = \exp(-i\frac{\phi}{2}\sigma_z \otimes \sigma_z)$ from single qubit rotations and the tunable CZ$_\phi$ entangling gate. To enable small and negative angles we include $\pi$ pulses around the X-axis ($A = X$) or Y-axis ($A = Y$). The unitary diagonals are $(1\ e^{i\phi}\ e^{i\phi}\ 1)$. (b) Tunable CZ$_\phi$ gate, implemented by moving $|ee\rangle$ (red) close to $|gf\rangle$ (blue). Coupling strength is $g/2\pi = 14$ MHz, pulse length is 55 ns, and typically $\Delta/2\pi = 0.7$ GHz when idling. (c) Measured versus desired phase of the full sequence, determined using quantum state tomography. (d) Quantum process tomography of anticommutation. The process matrices are shown for the non-trivial Hermitian terms of the anticommutation relations (left and center) as well as the sequence of both processes, yielding an identity unitary (right). The significant matrix elements (red and blue) are close to the ideal (transparent).

We now discuss the simulation of fermionic models. We use the Trotter approximation [25] to digitize the evolution of Hamiltonian $H = \sum_k H_k$: $U = \exp(-iHt) \simeq [\exp(-iH_1 t/n)\exp(-iH_2 t/n)...]^n$, with each part implemented using single- and two-qubit gates ($\hbar = 1$). We benchmark the simulation by comparing the experimental results to the exact digital outcome. Discretization unavoidably leads to deviations, and the digital errors are quantified in the Supplementary Information.

We start by visualizing the kinetic interactions between two fermionic modes. The construction of the Trotter step is shown in Fig. 3a and directly follows from the Hamiltonian in Eq. 2. The step consists of the $\hat{X}\hat{X}$, $\hat{Y}\hat{Y}$ and $\hat{Z}\hat{Z}$ terms, constructed from $\hat{Z}\hat{Z}$ terms and single qubit ro-tations. We simulate the evolution during time $\Delta t$ by setting $\phi_{xx} = \phi_{yy} = V\Delta t$ and $\phi_z = \phi_{zz} = U\Delta t/2$, and using $V = U = 1$. We evolve the system to a time of $T = 5.0$, and increase the number of steps ($\Delta t = T/n$, with $n = 1, ..., 8$). The data show hallmark oscillations, Fig. 3b, indicating that the modes interact and exchange excitations. We find that the end state fidelity [26], taken at the same simulated time, decreases approximately linearly by 0.054 per step (Fig. 3c).

The above example shows that fermionic simulations, clearly capturing the dynamics arising from interactions, can be performed digitally using single qubit gates and the tunable $CZ_\phi$ gate. Moreover, increasing the number of steps improves the time resolution, but at the price of increasing errors. A crucial result is that the per-step decrease in the end state fidelity is consistent with the gate fidelities. Using the typical values of $7.4 \cdot 10^{-3}$ entangling gate error and $8 \cdot 10^{-4}$ single qubit gate error as previously determined for this platform [9], we arrive at an expected Trotter step process error of 0.07, considering the step consists of six entangling gates and 28 single qubit gates (including X, Y rotations as well as idles). In addition, we have determined the Trotter step gate error in a separate interleaved randomized benchmarking experiment (Supplementary Information), and found a process error of 0.074, which is consistent with the observed per-step state error. We find that the process fidelity is thus a useful estimate, even though the simulation fidelity depends on the state and implemented model.

Simulations of fermionic models with three and four modes are shown in Fig. 4. The three-mode Trotter step and its pulse-sequence are shown in Figs. 4a-b. An implementation of the $\hat{Y}\hat{Y}$ gate is highlighted: the top qubit (red) is passive and detuned away, the middle qubit (blue) is tuned to an optimal fre-quency for the interaction, and the bottom qubit (green) performs the adiabatic trajectory. $\pi$-pulses on the passive qubit suppress dephasing and parasitic interactions. Fig. 4c shows the simulation results for $V = 1$, $U = 0$ (hopping only) and $V = 1$, $U = 1$ (with on-site repulsion). Input state generation is shown in the Supplementary Information. The simulation data (closed symbols) follows the exact digital outcome (open symbols), accumulating a per-step error of 0.15 (Fig. 4f) and gradually populating other states (black symbols). The fidelity is the relevant figure of merit, especially for simulations with few steps; the per-step error being the same for different models indicates that the simulated time evolutions are distinct.

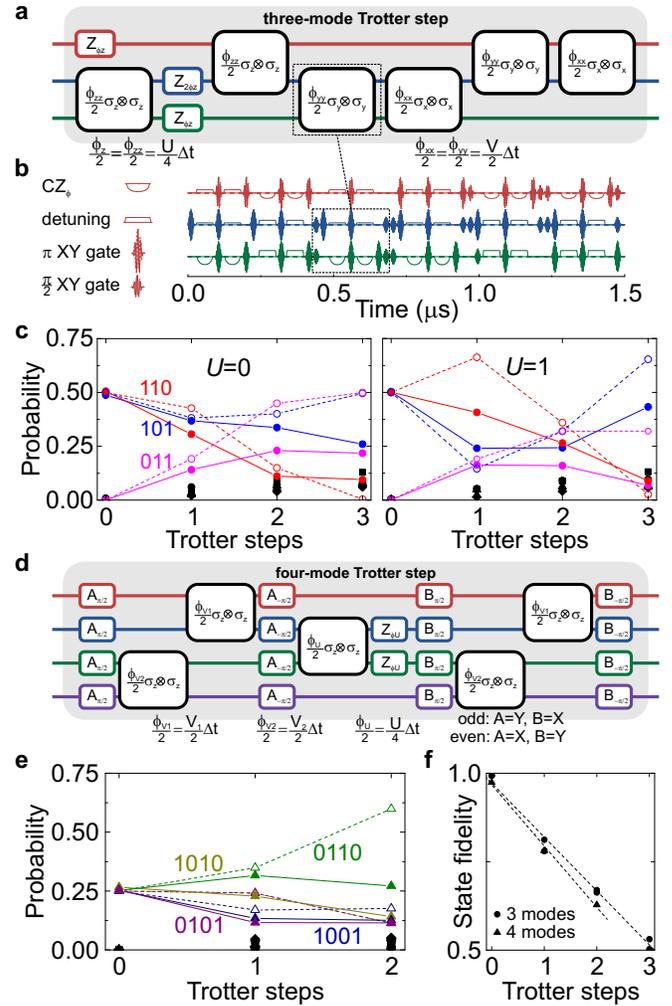

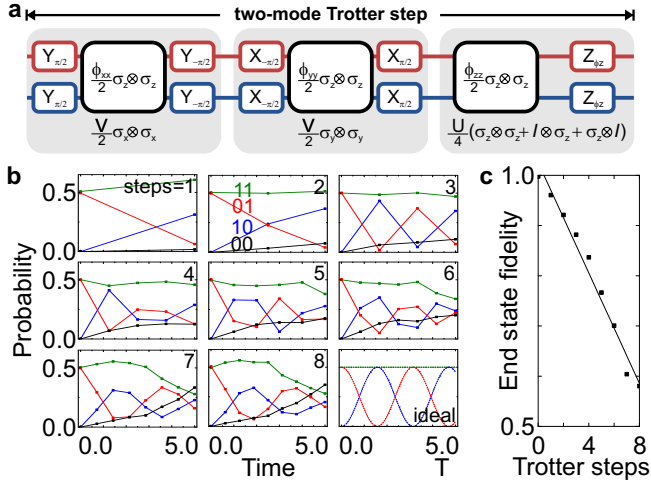

FIG. 3. **Simulation of two fermionic modes.** (a) Construction of the two-mode Trotter step, showing the separate terms of the Hamiltonian (Eq. 2). See Supplementary Information for the pulse sequence and gate count. (b) Probability of the modes versus simulated time for $n = 1, ..., 8$ steps. Input state is $[(|0\rangle + |1\rangle) \otimes |1\rangle]/\sqrt{2}$, and $V = U = 1$. The ideal dependence is shown in the bottom right. (c) The end-state fidelity decreases with step by 0.054, following a linear trend.

FIG. 4. **Fermionic models with three and four modes.** (a) Three-mode Trotter step, with the Trotter step pulse-sequence in (b). The Trotter step consists of 12 entangling gates and 87 single qubit gates (see text). The $\hat{Y}\hat{Y}$ interaction is highlighted (dashed). (c) Simulation results for three modes with and without on-site interaction. Full symbols: experiment. Open symbols: ideal digitized. Black symbols: population of other states. Input state is $[|1\rangle \otimes (|01\rangle + |10\rangle)]/\sqrt{2}$, and $V = 1$. (d) Construction of the four mode Trotter step. (e) Four mode simulation results for $V_1 = V_2 = 1$, $U_{23} = 1$, and $U_{14} = 0$. Input state is $[(|01\rangle + |10\rangle) \otimes (|01\rangle + |10\rangle)]/2$. (f) Simulation fidelities.

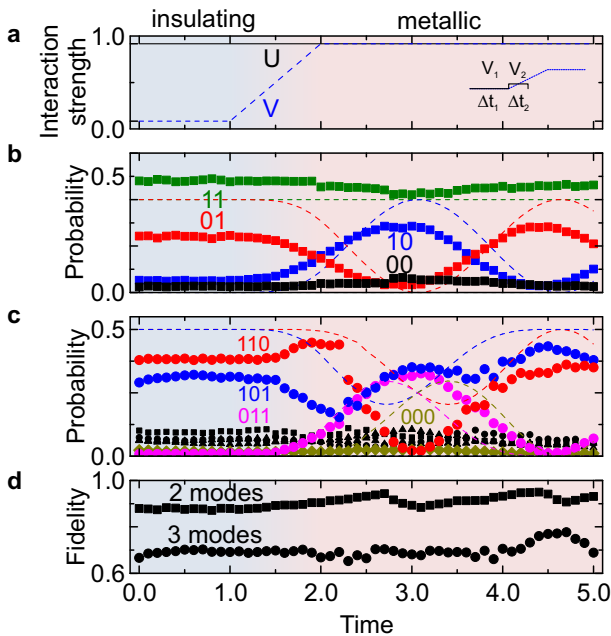

FIG. 5. **Simulations with time-varying interactions.** (a) The system is changed from an insulating state to a conducting phase, by ramping the hopping term $V$ from zero to one. Inset shows the choice of digitization on the ramp for the two-mode simulation. (b) Two-mode simulation showing dynamic behaviour starting at the onset of the $V$ ramp. (c) Three mode simulation, showing non-trivial dynamics when the hopping term is nonzero. (d) Simulation fidelities.

For the four-mode experiment, we simulate an asymmetric variation on the Hubbard model. Here, the repulsive interaction is between the middle modes only (right well in Fig. 1a), while the hopping terms are kept equal. Asymmetric models are used in describing anisotropic fermionic systems [27]. In addition, the simulation can be optimized: gate count is reduced by the removal of interaction between the top and bottom modes, and the Trotter expansion can be rewritten in terms of odd and even steps such that the starting and ending single-qubit gates cancel (Supplementary Information). The Trotter step is shown in Fig. 4d. The results are plotted in Fig. 4e. We find that the state fidelity decreases by 0.17 for the four mode simulation, see Fig. 4f.

The three- and four-mode experiments underline that fermionic models can be simulated digitally with large numbers of gates. The three-mode simulation uses in excess of 300 gates. We perform three Trotter steps, and per step we use: 12 entangling gates, 53 microwave $\pi$ and $\pi/2$ gates, 19 idle gates, 3 single-qubit phase gates, and for the non-participating qubit during the entangling operation: 12 frequency detuning gates where phases need to be accurately tracked. Using the above typical errors for gates, we arrive at an estimated process error of 0.16 for the three mode simulation, and an error of 0.15 for the four-mode simulation (per four mode Trotter step: 10 entangling gates and 98 single qubit gates). The process errors are close to the observed drop in state fidelity. Importantly, these results strongly suggest that the simulation errors scale with the number of gates, not qubits (modes), which is a crucial aspect of scalably implementing models on our platform.

We now address the simulation of fermionic systems with time-dependent interactions. In Fig. 5a, we show an experiment where we ramp the hopping term $V$ from 0 to 1 while keeping the on-site repulsion $U$ at 1; essentially changing the system from an insulating to a metallic phase. This transition is simulated for two modes using two Trotter steps, see inset, and with one step for three modes. For the latter case, we take the average of $V$ over the relevant time domain. The data are shown in Figs. 5b-c, and clearly mirror the dynamics of the hopping term. At time smaller than 1.0, the system is frozen and the mode probabilities are virtually unchanged, reflecting the insulating state. Interactions become visible when hopping is turned on, effectively melting the system, and follow the generic features of the exact digital outcome (dashed). The simulation fidelities lie around 0.9-0.95 for two modes and 0.7-0.8 for three modes, see Fig. 5d. These fidelities are around or somewhat below those for time evolution with constant interactions, presumably due to control errors related to parasitic qubit interactions, which also lead to the populating of other states (black symbols). The dynamic simulation highlights the possibilities of exploring parameter spaces and transitions with few steps.

We have demonstrated the digital quantum simulation of fermionic models. Simulation fidelities are close to the expected values, and with improvements in gates and architecture, the construction of larger testbeds for fermionic systems appears viable. Bosonic modes can be elegantly introduced by adding linear resonators to the circuit, establishing a fermion-boson analog-digital system [15, 16] as a distinct paradigm for quantum simulation.

**Methods Summary** Experiments are performed in a wet dilution refrigerator with a base temperature of 20 mK. Qubit frequencies are chosen in a staggered pattern to minimize unwanted interaction. Typical qubit frequencies are 5.5 and 4.8 GHz. Exact frequencies are optimized based on the qubits' $|e\rangle$ and $|f\rangle$ state spectra along the fully tunable trajectory of the $CZ_\phi$-gate, as well as minimizing interaction between next-nearest neighbouring qubits. Used qubits are Q1-Q4 in Ref. [19]. Data are corrected for measurement fidelity.

**Acknowledgements** We thank A. N. Korotkov for discussions. The authors acknowledge support from Spanish MINECO FIS2012-36673-C03-02; Ramón y Cajal Grant RYC-2012-11391; UPV/EHU UFI 11/55 and EHUA14/04; Basque Government IT472-10; a UPV/EHU PhD grant; PROMISCE and SCALEQIT EU projects. Devices were made at the UC Santa Barbara Nanofabrication Facility, a part of the NSF-funded National Nanotechnology Infrastructure Network, and at the NanoStructures Cleanroom Facility.

**Author Contributions** R.B., L.L., and L.G.-Á. designed the experiment, with J.M.M. and E.S. providing supervision. L.G.-Á., L.L., and E.S. provided the theoretical framework. R.B. and L.L. cowrote the manuscript with J.M.M. and E.S. The experiment and data were performed and analysed by R.B., J.K., L.L., and L.G.-Á. R.B. and J.K. designed the device. J.K., R.B., and A.M. fabricated the sample. All authors contributed to the fabrication process, experimental set-up and manuscript preparation.

# Supplementary Information for "Digital quantum simulation of fermionic models with a superconducting circuit"

R. Barends,[1] L. Lamata,[2] J. Kelly,[3] L. García-Álvarez,[2] A. G. Fowler,[1] A. Megrant,[3,4] E. Jeffrey,[1] T. C. White,[3] D. Sank,[1] J. Y. Mutus,[1] B. Campbell,[3] Yu Chen,[1] Z. Chen,[3] B. Chiaro,[3] A. Dunsworth,[3] I.-C. Hoi,[3] C. Neill,[3] P. J. J. O'Malley,[3] C. Quintana,[3] P. Roushan,[1] A. Vainsencher,[3] J. Wenner,[3] E. Solano,[2,5] and John M. Martinis[1,3]

[1]*Google Inc., Santa Barbara, CA 93117, USA*
[2]*Department of Physical Chemistry, University of the Basque Country UPV/EHU, Apartado 644, E-48080 Bilbao, Spain*
[3]*Department of Physics, University of California, Santa Barbara, CA 93106, USA*
[4]*Department of Materials, University of California, Santa Barbara, CA 93106, USA*
[5]*IKERBASQUE, Basque Foundation for Science, Maria Diaz de Haro 3, 48013 Bilbao, Spain.*

## I. TROTTER STEP PULSE SEQUENCES

### A. Pulse sequences and gate counts

The two-, three- and four-mode Trotter step pulse sequences are shown in Fig. S1. The gate counts can be found in Table S1.

### B. Initialization

The gate sequences for the initialization of the three- and four-mode simulation are shown in Fig. S2. For the two-mode simulation the input state is: $[(|0\rangle + |1\rangle) \otimes |1\rangle]/\sqrt{2}$, for three modes: $[|1\rangle \otimes (|01\rangle + |10\rangle)]/\sqrt{2}$, and for four modes: $[(|01\rangle + |10\rangle) \otimes (|01\rangle + |10\rangle)]/2$.

## II. QUANTUM PROCESS TOMOGRAPHY

We use quantum process tomography to determine the $\chi$ matrix. We start by initializing the qubits into the ground state, and prepare input states by applying gates from $\{I, X/2, Y/2, X\}^{\otimes 2}$. The process output is reconstructed by applying gates from the same group, essentially obtaining the 16 output density matrices. The $\chi$ matrix is then determined using quadratic maximum likelihood estimation, using the MATLAB packages SeDuMi and YALMIP, while constraining it to be Hermitian, trace-preserving, and positive semidefinite; the estimation is overconstrained. Non-idealities in measurement and state preparation are suppressed by performing tomography on a zero-time idle.

The $\chi$ matrices for processes $U_1 = \exp(-i\frac{\pi}{2}(b_1 b_2^\dagger + b_2 b_1^\dagger))$ and $U_2 = \exp(-i\frac{\pi}{2}(b_1^\dagger b_2 + b_2^\dagger b_1))$ are determined experimen-tally, and the matrix of process $U_2 U_1$ is computed from the experimentally obtained matrices following Ref. [1].

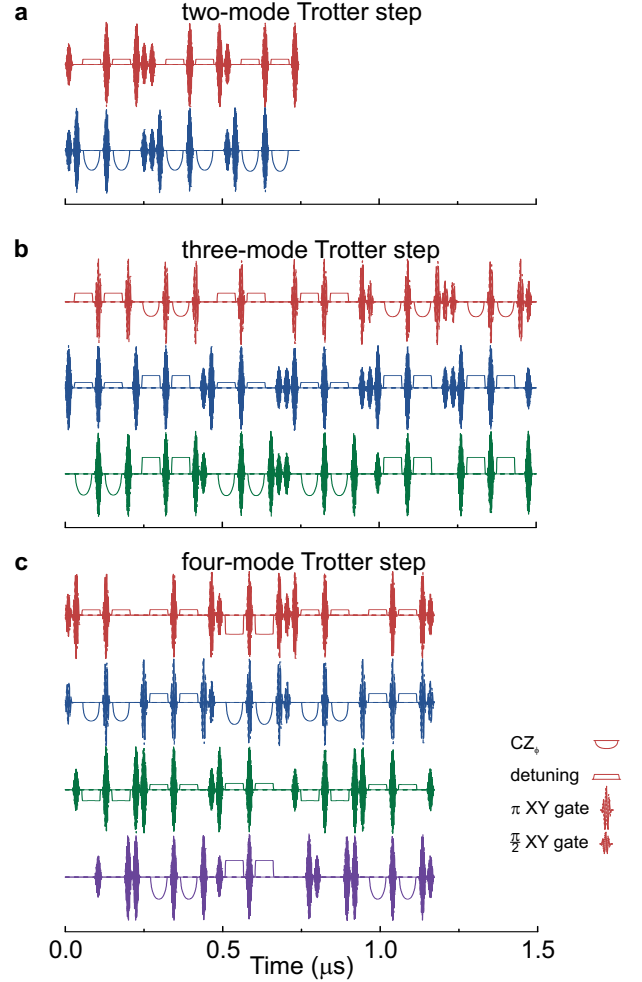

FIG. S1. Pulse sequences for a single two-mode (a), three-mode (b), and four-mode (c) Trotter step. Shown are entangling gates as well as single-qubit microwave, idle and detuning gates. The legend is in the bottom right.

TABLE S1. Gate counts for the two-, three-, and four-mode Trotter step, determined using Fig. S1. We count idles as having the same duration as the microwave $\pi$ and $\pi/2$ gates; this is the relevant approach for estimating total process fidelities. The gate counts are for a single Trotter step only, and exclude input state preparation.

| Gates | two-mode | three-mode | four-mode |
|---|---|---|---|
| entangling $CZ_\phi$ | 6 | 12 | 10 |
| single qubit | 28 | 87 | 98 |
| - microwave $\pi$ and $\pi/2$ | 20 | 53 | 56 |
| - idle | 6 | 19 | 22 |
| - detuning | 0 | 12 | 18 |
| - virtual phase | 2 | 3 | 2 |

The used quantum circuits are

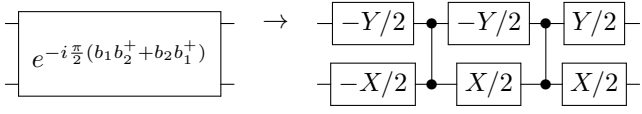

and

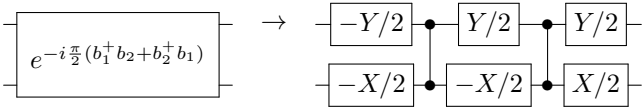

## III. RANDOMIZED BENCHMARKING OF $\exp(-i\frac{\pi}{4}\sigma_z \otimes \sigma_z)$ AND THE TWO-MODE TROTTER STEP

The process fidelity of the $\exp(-i\frac{\phi}{2}\sigma_z \otimes \sigma_z)$ gate and the two-mode Trotter step are determined using interleaved Clifford-based randomized benchmarking [2–4]. This technique is insensitive to measurement and state preparation error, and determines the fidelity properly averaged over all input states, but it restricts the gates to have a unitary which lies within the group of Cliffords. As representative angles we have therefore used $\phi = \pi/2$, and $\phi_{xx} = \phi_{yy} = \phi_{zz} = \pi/2$ for the Trotter step.

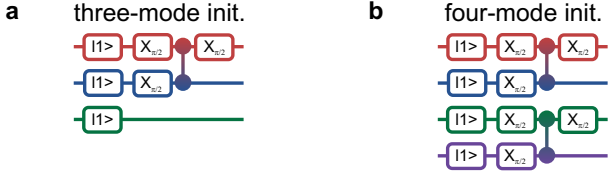

FIG. S2. Initialization gate sequence. (a) Three-mode initialization. (b) Four-mode initialization.

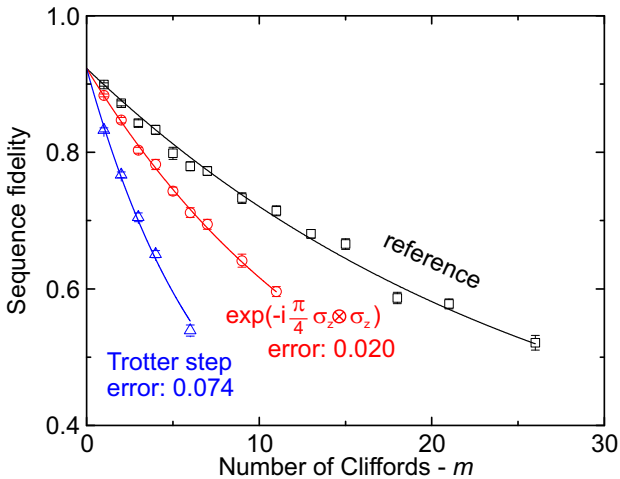

FIG. S3. Clifford-based randomized benchmarking of $\exp(-i\frac{\pi}{4}\sigma_z \otimes \sigma_z)$ and the two-mode Trotter step. Sequence fidelity versus number of Cliffords. Black: reference. Colour: interleaved.

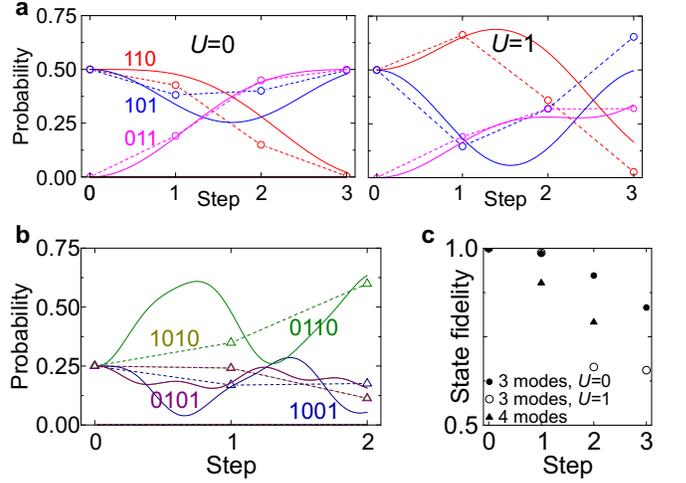

FIG. S4. Digital error for the time-independent simulation. (a) Three mode simulation ($U = 0, U = 1, V = 1$). (b) Four mode simulation ($U_{23} = 1, U_{14} = 0, V = 1$). (c) Fidelity. Ideal evolution (solid lines) and exact digital solution (open symbols connected by dashed lines).

The data are shown in Fig. S3. We start by measuring the decay in sequence fidelity of sequences of random, two-qubit Cliffords (black symbols). When interleaving we see an extra decrease of sequence fidelity, which can be linked to the process fidelity of the interleaved gate. We find that the $\exp(-i\frac{\pi}{4}\sigma_z \otimes \sigma_z)$ gate and the Trotter step have errors of 0.020 and 0.074, respectively. We note that these values are consistent with estimation by adding individual gate errors (main Letter).

## IV. DIGITAL ERROR

The Trotter expansion introduces digital errors due to discretization. A full analysis of the digital error for the used model can be found in Ref. [5]. For the time-independent model, the two-mode simulation has zero digital error. For the three- and four-mode simulation the full evolution (solid lines), exact digital solution (open symbols connected by dashed lines), and fidelities due to digital error are shown in Fig. S4.

For the time-dependent model we find a negligible digital error for two modes, and a significant error for three, see Fig. S5. The large error for three modes arises from having to approximate a larger Hamiltonian, as well as using only a single step.

## V. MINIMIZING LEAKAGE OF THE CZ$_\phi$ GATE

The tunable CZ$_\phi$ gate works by tuning the frequency of one of the qubits to approach the avoided level crossing of the $|ee\rangle$ and $|gf\rangle$ states, using an adiabatic trajectory [6]. For large phases we need to closely approach the avoided level crossing, inducing state leakage.

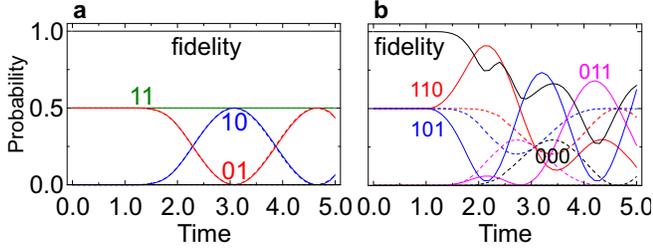

FIG. S5. Digital error for the time-dependent simulation for two modes, using two Trotter steps (a) and three modes, using one Trotter step (b). Ideal evolution (solid lines), exact digital solution (dashed lines), and fidelity (solid black).

To minimize such leakage we have chosen to increase the length of the $CZ_\phi$ gate from a typical 40 ns [7] to 55 ns. However, for large phases ($> 4.0$ rads), see Fig. S6a, we still see a considerable amount of leakage, see the Fig. S6b. By choosing the leaked state population as a fitness metric, and using Nelder-Mead optimization in a similar approach to Ref. [8] to tune waveform parameters, see Figs. S6c-d, we can significantly suppress leakage. We note that this optimization took approximately one minute in real time.

## VI. ASYMMETRIC HUBBARD MODEL

Here, we include the analysis of the fermionic asymmetric Hubbard model for 4 qubits employed in the Letter. Firstly, we present the model in terms of spin operators via the Jordan-Wigner transformation, and describe different limits of the model. Secondly, we analyse the digital quantum simulation in terms of Trotter steps involving the optimized gates ($CZ_\phi$).

The asymmetric Hubbard model (AHM) is a variation of the Hubbard model that describes anisotropic fermionic systems. Here, we are going to consider this model for two different fermionic species, that could represent spins, interacting with each other by the Coulomb term, and two lattice sites. The operators for this model have two indices, $A_{ij}$, where $i$ and $j$ indicate the site position and kind of particle, respectively. Since the fermions might have different masses, we have no reason to assume that the hopping terms will be the same. We can write the Hamiltonian for two sites, $x$ and $y$, and two kinds of fermions, 1 and 2, as

$$H = - V_1 \left( b^\dagger_{x1} b_{y1} + b^\dagger_{y1} b_{x1} \right)$$
$$- V_2 \left( b^\dagger_{x2} b_{y2} + b^\dagger_{y2} b_{x2} \right)$$
$$+ U_x b^\dagger_{x1} b_{x1} b^\dagger_{x2} b_{x2}$$
$$+ U_y b^\dagger_{y1} b_{y1} b^\dagger_{y2} b_{y2}, \quad (S1)$$

where $b^\dagger_{mi}$ and $b_{mi}$ are fermionic creation and annihilation operators of the kind of particle $i$ for the site $m$. For the main Letter we use $b^\dagger_1, b^\dagger_2, b^\dagger_3, b^\dagger_4$, for $b^\dagger_{x1}, b^\dagger_{y1}, b^\dagger_{y2}, b^\dagger_{x2}$.

The Jordan-Wigner transformation will be used in our derivation to relate the fermionic and antifermionic operators with tensor products of Pauli matrices, which are operators that we can simulate in the superconducting circuit setup.

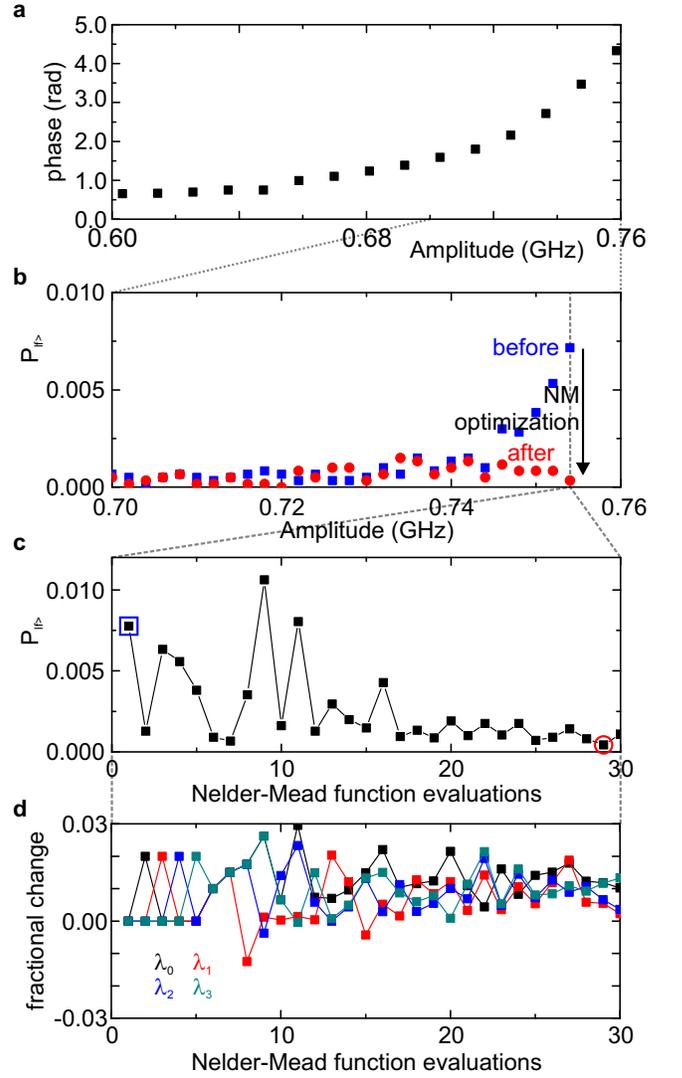

FIG. S6. Minimizing leakage of the $CZ_\phi$ gate. (a) Tunable phase versus pulse amplitude, determined with quantum state tomography. (b) Zoom-in of the amplitude region for large phases, showing the $|f\rangle$-state population before (blue) and after (red) Nelder-Mead optimization. (c) Population of $|f\rangle$ versus Nelder-Mead function evaluation, showing a downwards trend. (d) Optimization of the waveform parameters with Nelder-Mead function evaluation, see Ref. [6] for the definition of these parameters.

This transformation is based on a mapping between fermionic operators and spin-1/2 operators. In this case, the relations are

$$b^\dagger_{x1} = \mathbb{I} \otimes \mathbb{I} \otimes \mathbb{I} \otimes \sigma^+$$
$$b^\dagger_{y1} = \mathbb{I} \otimes \mathbb{I} \otimes \sigma^+ \otimes \sigma^z$$
$$b^\dagger_{y2} = \mathbb{I} \otimes \sigma^+ \otimes \sigma^z \otimes \sigma^z$$
$$b^\dagger_{x2} = \sigma^+ \otimes \sigma^z \otimes \sigma^z \otimes \sigma^z. \quad (S2)$$

After this mapping, Hamiltonian (S1) is rewritten in terms

of spin-1/2 operators as

$$H = \frac{V_1}{2}\left(\mathbb{I}\otimes\mathbb{I}\otimes\sigma^x\otimes\sigma^x + \mathbb{I}\otimes\mathbb{I}\otimes\sigma^y\otimes\sigma^y\right) + \frac{V_2}{2}\left(\sigma^x\otimes\sigma^x\otimes\mathbb{I}\otimes\mathbb{I} + \sigma^y\otimes\sigma^y\otimes\mathbb{I}\otimes\mathbb{I}\right)$$
$$+ \frac{U_x}{4}\left(\sigma^z\otimes\mathbb{I}\otimes\mathbb{I}\otimes\sigma^z + \mathbb{I}\otimes\mathbb{I}\otimes\mathbb{I}\otimes\sigma^z + \sigma^z\otimes\mathbb{I}\otimes\mathbb{I}\otimes\mathbb{I}\right)$$
$$+ \frac{U_y}{4}\left(\mathbb{I}\otimes\sigma^z\otimes\sigma^z\otimes\mathbb{I} + \mathbb{I}\otimes\mathbb{I}\otimes\sigma^z\otimes\mathbb{I} + \mathbb{I}\otimes\sigma^z\otimes\mathbb{I}\otimes\mathbb{I}\right), \tag{S3}$$

where the different interactions can be simulated via digital techniques in terms of single qubit and $CZ_\phi$ gates.

### A. Gate decomposition

We consider the digital quantum simulation of the dynamics of Hamiltonian (S3). The Trotter expansion consists of dividing the time $t$ into $n$ time intervals of length $t/n$, and applying sequentially the evolution operator of each term of the Hamiltonian for each time interval. In this case the evolution operators are associated with the different summands of the Hamiltonian.

In order to describe the digital simulation in terms of Trotter steps involving the optimized gates ($CZ_\phi$), we will first consider the Hamiltonian in terms of $\exp[-i(\phi/2)\sigma^z\otimes\sigma^z]$ interactions. We take into account the relations

$$\sigma^x\otimes\sigma^x = R_y(\pi/2)\sigma^z\otimes\sigma^z R_y(-\pi/2)$$
$$\sigma^y\otimes\sigma^y = R_x(-\pi/2)\sigma^z\otimes\sigma^z R_x(\pi/2), \tag{S4}$$

where $R_j(\theta) = \exp(-i\frac{\theta}{2}\sigma^j)$ is the rotation along the $j$ coordinate of a qubit. In these expressions the rotations are applied on the two qubits of the product.

The evolution operator associated with Hamiltonian (S3) in terms of $\exp[-i(\phi/2)\sigma^z\otimes\sigma^z]$ interactions is

$$e^{-iHt} \approx \prod_k \left(e^{-iH_k\frac{t}{n}}\right)^n$$
$$\approx \Bigg(R_y(\pi/2)e^{-i\frac{V_1}{2}\mathbb{I}\otimes\mathbb{I}\otimes\sigma^z\otimes\sigma^z\frac{t}{n}}R_y(-\pi/2)R_x(-\pi/2)e^{-i\frac{V_1}{2}\mathbb{I}\otimes\mathbb{I}\otimes\sigma^z\otimes\sigma^z\frac{t}{n}}R_x(\pi/2)$$
$$\cdot R_y(\pi/2)e^{-i\frac{V_2}{2}\sigma^z\otimes\sigma^z\otimes\mathbb{I}\otimes\mathbb{I}\frac{t}{n}}R_y(-\pi/2)R_x(-\pi/2)e^{-i\frac{V_2}{2}\sigma^z\otimes\sigma^z\otimes\mathbb{I}\otimes\mathbb{I}\frac{t}{n}}R_x(\pi/2)$$
$$\cdot e^{-i\frac{U_x}{4}\sigma^z\otimes\mathbb{I}\otimes\mathbb{I}\otimes\sigma^z\frac{t}{n}}e^{-i\frac{U_x}{4}\mathbb{I}\otimes\mathbb{I}\otimes\mathbb{I}\otimes\sigma^z\frac{t}{n}}e^{-i\frac{U_x}{4}\sigma^z\otimes\mathbb{I}\otimes\mathbb{I}\otimes\mathbb{I}\frac{t}{n}}$$
$$\cdot e^{-i\frac{U_y}{4}\mathbb{I}\otimes\sigma^z\otimes\sigma^z\otimes\mathbb{I}\frac{t}{n}}e^{-i\frac{U_y}{4}\mathbb{I}\otimes\mathbb{I}\otimes\sigma^z\otimes\mathbb{I}\frac{t}{n}}e^{-i\frac{U_y}{4}\mathbb{I}\otimes\sigma^z\otimes\mathbb{I}\otimes\mathbb{I}\frac{t}{n}}\Bigg)^n. \tag{S5}$$

Note that, in principle, the ordering of the gates inside a Trotter step does not have a sizable effect as far as there are enough Trotter steps. Here, the number of Trotter steps is limited ($n$ approximately $\leq 10$) and different orderings will have different results. The different values in the orderings differ in a $O(1)$ constant, while the global digital error depends on the number of Trotter steps $n$ as $1/n$ (the difference in errors due to different orderings does not depend on $n$).

If we consider the Trotter error, the fidelity could increase with an optimal ordering where we group terms of the Hamiltonian that commute with each other. Nevertheless, from the experimental point of view, the operators can be rearranged in a more suitable way in order to optimize the number of gates and eliminate global phases. In this sense, we must look for the optimal ordering by considering both aspects.

Here, we simply rearrange the operators in order to optimize the number of gates. If we consider that $R_j(\alpha) + R_j(\beta) = R_j(\alpha+\beta)$, then

$$e^{-iHt} \approx \prod_{i=1}^{n/2} \Bigg( R'_y(\pi/2) e^{-i\frac{V_1}{2} \mathbb{I} \otimes \mathbb{I} \otimes \sigma^z \otimes \sigma^z \frac{t}{n}} R'_y(-\pi/2) R_y(\pi/2) e^{-i\frac{V_2}{2} \sigma^z \otimes \sigma^z \otimes \mathbb{I} \otimes \mathbb{I} \frac{t}{n}} R_y(-\pi/2)$$
$$\cdot e^{-i\frac{U_x}{4} \sigma^z \otimes \mathbb{I} \otimes \mathbb{I} \otimes \sigma^z \frac{t}{n}} e^{-i\frac{U_x}{4} \mathbb{I} \otimes \mathbb{I} \otimes \mathbb{I} \otimes \sigma^z \frac{t}{n}} e^{-i\frac{U_x}{4} \sigma^z \otimes \mathbb{I} \otimes \mathbb{I} \otimes \mathbb{I} \frac{t}{n}}$$
$$\cdot e^{-i\frac{U_y}{4} \mathbb{I} \otimes \sigma^z \otimes \sigma^z \otimes \mathbb{I} \frac{t}{n}} e^{-i\frac{U_y}{4} \mathbb{I} \otimes \mathbb{I} \otimes \sigma^z \otimes \mathbb{I} \frac{t}{n}} e^{-i\frac{U_y}{4} \mathbb{I} \otimes \sigma^z \otimes \mathbb{I} \otimes \mathbb{I} \frac{t}{n}}$$
$$\cdot R'_x(-\pi/2) e^{-i\frac{V_1}{2} \mathbb{I} \otimes \mathbb{I} \otimes \sigma^z \otimes \sigma^z \frac{t}{n}} R'_x(\pi/2) R_x(-\pi/2) e^{-i\frac{V_2}{2} \sigma^z \otimes \sigma^z \otimes \mathbb{I} \otimes \mathbb{I} \frac{t}{n}} R_x(\pi/2) \Bigg)_{2i-1}$$
$$\cdot \Bigg( R'_x(-\pi/2) e^{-i\frac{V_1}{2} \mathbb{I} \otimes \mathbb{I} \otimes \sigma^z \otimes \sigma^z \frac{t}{n}} R'_x(\pi/2) R_x(-\pi/2) e^{-i\frac{V_2}{2} \sigma^z \otimes \sigma^z \otimes \mathbb{I} \otimes \mathbb{I} \frac{t}{n}} R_x(\pi/2)$$
$$\cdot e^{-i\frac{U_x}{4} \sigma^z \otimes \mathbb{I} \otimes \mathbb{I} \otimes \sigma^z \frac{t}{n}} e^{-i\frac{U_x}{4} \mathbb{I} \otimes \mathbb{I} \otimes \mathbb{I} \otimes \sigma^z \frac{t}{n}} e^{-i\frac{U_x}{4} \sigma^z \otimes \mathbb{I} \otimes \mathbb{I} \otimes \mathbb{I} \frac{t}{n}}$$
$$\cdot e^{-i\frac{U_y}{4} \mathbb{I} \otimes \sigma^z \otimes \sigma^z \otimes \mathbb{I} \frac{t}{n}} e^{-i\frac{U_y}{4} \mathbb{I} \otimes \mathbb{I} \otimes \sigma^z \otimes \mathbb{I} \frac{t}{n}} e^{-i\frac{U_y}{4} \mathbb{I} \otimes \sigma^z \otimes \mathbb{I} \otimes \mathbb{I} \frac{t}{n}}$$
$$\cdot R'_y(\pi/2) e^{-i\frac{V_1}{2} \mathbb{I} \otimes \mathbb{I} \otimes \sigma^z \otimes \sigma^z \frac{t}{n}} R'_y(-\pi/2) R_y(\pi/2) e^{-i\frac{V_2}{2} \sigma^z \otimes \sigma^z \otimes \mathbb{I} \otimes \mathbb{I} \frac{t}{n}} R_y(-\pi/2) \Bigg)_{2i}, \tag{S6}$$

where we use the prime notation in the rotation to distinguish between gates applied on different qubits. This decomposition between even and odd Trotter steps is suitable in order to simplify rotations in $x$ and $y$, and, therefore, avoid higher number of gates.

The sequence of gates for one odd Trotter step in the digital simulation of the Hubbard model with four qubits is

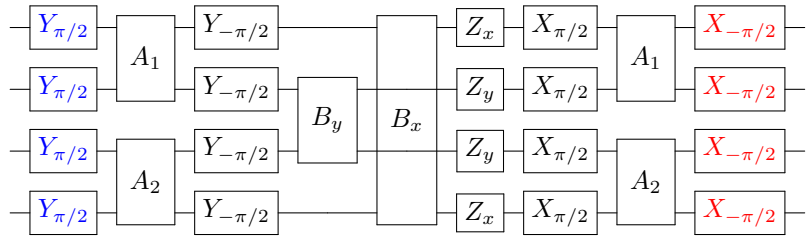

and for one even Trotter step:

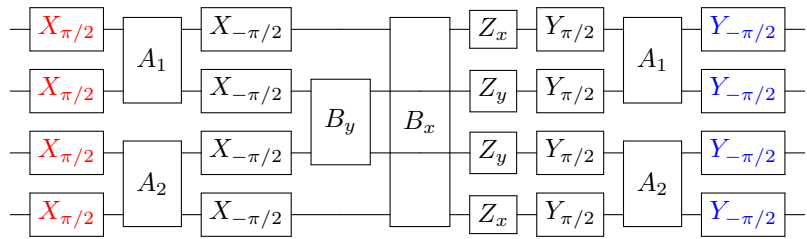

The gates $A_i$ and $B_j$ are two-qubit gates in terms of the $\exp[-i(\phi/2)\sigma^z \otimes \sigma^z]$ interactions: $A_i = \exp(i\frac{V_i}{2}\sigma^z \otimes \sigma^z \frac{t}{n})$ and $B_j = \exp(-i\frac{U_j}{4}\sigma^z \otimes \sigma^z \frac{t}{n})$. The $Z_i$ gates are single qubit rotations: $Z_i = \exp(-i\frac{U_i}{4}\sigma^z \frac{t}{n})$, and $X_\alpha$ and $Y_\alpha$ are rotations along the $x$ and $y$ axis, respectively.

The $\exp[-i(\phi/2)\sigma^z \otimes \sigma^z]$ interaction can be implemented in small steps with optimized $CZ_\phi$ gates. The interaction is

$$e^{-i\frac{\phi}{2}\sigma^z \otimes \sigma^z} = \begin{pmatrix} 1 & & & \\ & e^{i\phi} & & \\ & & e^{i\phi} & \\ & & & 1 \end{pmatrix}.$$

The quantum circuits for simulating this are shown in the main Letter.

### B. Particular case of the model

In order to avoid the gate $B_x$ between the first and the fourth qubit, we can consider a particular case of the asymmetric Hubbard model, where $U_x = 0$. In this case, the circuit is the same but without the $B_x$ and the $Z_x$ gates. That is, for one odd Trotter step:

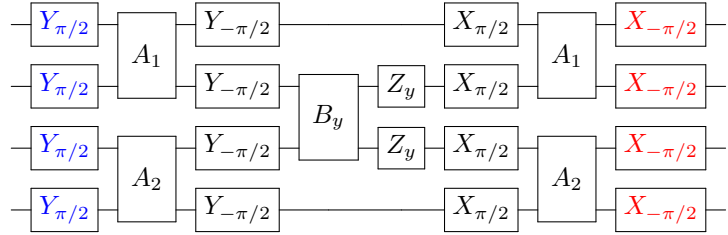

and for one even Trotter step:

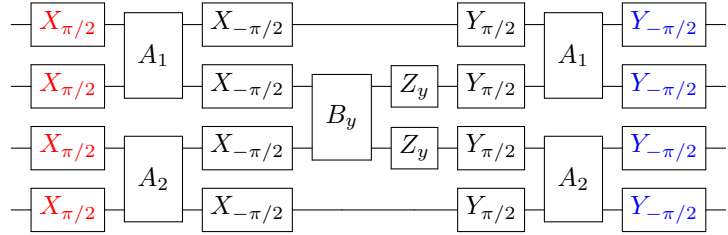

It is important to note that, for $n = 2$ Trotter steps, the red gates cancel each other, and we reduce the number of gates that should be applied. For $n > 2$, the blue gates also cancel each other except in the beginning and in the end of the quantum simulation.

### C. Digital quantum simulation of the model

The relation among the values of the parameters in the numerical simulations and the values of the phases in the gates is the following

$$A_1 = \exp(-i\frac{V_1}{2}\sigma^z \otimes \sigma^z \frac{t}{n}) \;\;\to\;\; \Phi_{A_1} = \frac{V_1}{2}\frac{t}{n}$$

$$A_2 = \exp(-i\frac{V_2}{2}\sigma^z \otimes \sigma^z \frac{t}{n}) \;\;\to\;\; \Phi_{A_2} = \frac{V_2}{2}\frac{t}{n}$$

$$B_y = \exp(-i\frac{U_y}{4}\sigma^z \otimes \sigma^z \frac{t}{n}) \;\;\to\;\; \Phi_{B_y} = \frac{U_y}{4}\frac{t}{n}$$

$$Z_y = \exp(-i\frac{U_y}{4}\sigma^z \frac{t}{n}) \;\;\to\;\; \Phi_2 = \frac{U_y}{4}\frac{t}{n} \;.$$

Notice that in the numerical simulations we consider $\hbar = 1$ for simplicity.

In summary, the fermionic asymmetric Hubbard model with two excitations, one for each kind of fermion, has been analysed and expressed in terms of simulatable spin operators. We have considered the digital quantum simulation in terms of Trotter steps involving the optimized gates (CZ$_\phi$). This is the four-mode system experimentally simulated in the main Letter.